\documentclass[12pt]{iopart}

\usepackage{iopams}
\usepackage{epsfig}
\usepackage{graphicx}
\usepackage{amsfonts}
\usepackage[figuresright]{rotating}
\usepackage{amssymb}
\usepackage{psfrag}

\def\avg#1{\langle#1\rangle}

\def\be{\begin{equation}}
\def\ee{\end{equation}}
\def\bea{\begin{eqnarray}}
\def\eea{\end{eqnarray}}

\def\nn{\nonumber}

\begin{document}

\title{Spontaneous breaking of time-reversal symmetry 
in the orbital channel for the boundary Majorana flat bands}

\author{Yi Li}
\address{Department of Physics, University of California, San Diego,
CA 92093, USA}
\ead{y4li@physics.ucsd.edu}
\author{Da Wang}
\address{Department of Physics, University of California, San Diego,
CA 92093, USA}
\ead{d6wang@ucsd.edu}
\author{Congjun Wu}
\address{Department of Physics, University of California, San Diego,
CA 92093, USA}
\ead{wucj@physics.ucsd.edu}

\begin{abstract}
We study the boundary Majorana modes for the single component $p$-wave weak
topological superconductors or superfluids, which form zero energy
flat bands protected by time-reversal symmetry in the orbital channel.
However, due to the divergence of density of states, the band flatness
of the edge Majorana modes
is unstable under spontaneously generated spatial variations of
Cooper pairing phases. 
Staggered current loops appear near the boundary and thus 
time-reversal symmetry is spontaneously broken in the orbital channel.
This effect can appear in both condensed matter and ultra-cold atom
systems.
\end{abstract}


\maketitle

\section{Introduction}
Recently, Majorana fermions in unconventional superconductors and pairing
superfluids have become a research focus in condensed matter physics
\cite{kopnin1991,volovik1999,read2000,ivanov2001,kitaev2001,
sengupta2001,kwon2004,stone2006,fu2008,fu2009,qi2009,cheng2009,chung2009,sau2010,sau2010a,mourik2012,churchill2013,das2012,finck2013,rokhinson2012}.
They appear at boundaries and in vortex cores exhibiting non-Abelian
statistics which potentially can be used for topological quantum 
computation \cite{ivanov2001, kitaev2001,teo2010,freedman2011,alicea2011,jiang2011,halperin2012}.
Various theoretical schemes have been proposed for the realization and
detection of Majorana fermions \cite{stanescu2013,alicea2012,
fu2008,sau2010,sau2010a,gurarie2007,zhangCW2008}, such as
the interface between 3D topological insulators and conventional
superconductors \cite{fu2008},
the proximity effect of superconductivity in spin-orbit coupled quantum
wires \cite{sau2010,sau2010a},
Cooper pairing with ultra-cold fermions with synthetic
spin-orbit coupling \cite{zhangCW2008},
and $p$-wave Feshbach resonances of single-component ultra-cold
fermions \cite{gurarie2007}.
Majorana fermions are also proposed in fractional quantum Hall states
at the filling $\nu=\frac{5}{2}$ \cite{dassarma2005,stern2006,nayak2008}.
The signature of the zero energy Majorana boundary mode
in the transport spectra has been observed in the spin-orbit coupled
quantum wires \cite{mourik2012,churchill2013,das2012,finck2013,rokhinson2012}.

The boundary Majorana zero modes also appear in 
unconventional superconductors in 2D and 3D
on edges or surfaces with suitable orientations determined by
Cooper pairing symmetries.
This is because along the directions of incident and reflected wavevectors
of the Bogoliubov quasi-particles, the pairing gap functions have opposite
signs \cite{hu1994,tanaka1995,kashiwaya2000}.
These unconventional superconductors can be viewed as weak topological
states.
They are topologically non-trivial (or trivial) along the directions
perpendicular to boundaries with (or without) zero energy modes,
respectively.

Interactions can significantly change topological properties of
edge and surface in topological insulators and superconductors.
For non-chiral systems, these boundary modes are low energy midgap
states, thus they are more susceptible to interactions than the
gapped bulk states.
In the helical edge Luttinger liquids of 2D topological insulators,
the edge magnetic fluctuations are much stronger than those in the
bulk, which can lead to spontaneous time-reversal symmetry
breaking and destroy edge modes \cite{wu2006,xu2006,zheng2011}.
For the zero energy boundary Majorana modes along the coupled chains
of the $p_x$-topological superconductors, it has been found that
interactions can even generate gaps without breaking time-reversal symmetry
in the orbital channel \cite{fidkowski2011}.
Similar effects are also found for the helical edge Majorana modes
of time-reversal invariant topological pairing states \cite{yao2012,qi2012}.
In both cases, the new topological class under interactions is
classified by a Z$_8$ periodicity.

In this article, we consider the most natural interaction effects in the
boundary states in the $p$-wave weak topological superconductors or
paired superfluids: the coupling between Cooper pairing phases and
the zero energy Majorana modes.
The degeneracy of these boundary modes is protected by time-reversal
symmetry, but is vulnerable under spontaneous time-reversal symmetry breaking
in the orbital channel.
Spatial variations of phases of Cooper pairing order parameters induce
bonding among these boundary modes, and thus lower the energy.
Staggered current loops are generated near the boundary,
which split the zero energy Majorana peaks.
Due to the divergence of the surface density of states, this time-reversal
symmetry breaking mechanism is robust.

The rest of this paper is organized as follows.
In Sect. \ref{sect:two-wires}, we explain the mechanism of 
spontaneous time-reversal symmetry breaking 
in the orbital channel wtih coupled 
superconducting quantum wires  through weak links.
In Sect. \ref{sect:boundary}, a 2D $p$-wave weak topological
superconductor is studied with the open boundary condition.
The spontaneous staggered orbital current loops appear on the
topological non-trivial boundaries,  and the consequential 
splitting of the Majorana zero bias peaks are calculated 
self-consistently.
Experimental realizations of the above effects are discussed in 
Sect. \ref{sect:discussions}.
The summary is presented in Sect. \ref{sect:conclusion}.

\section{Bonding between boundary Majorana modes}
\label{sect:two-wires}
We begin with a heuristic example of two parallel quantum wires along the
$x$-direction.
Each of them is a 1D topological superconductor of single component
fermion with $p_x$-pairing symmetry.
As explained in Ref. \cite{kitaev2001},  zero energy Majorana modes exist
near the two ends of each wire.
For example, for the end at $x=L$ of the $i$-th wire $(i=1,2)$,
the operator for the Majorana mode can be expressed as
\bea
\gamma_{i}=\int dx  \{ u_0(x) e^{-i\frac{\theta_i}{2}-i\frac{\pi}{4}}
\psi_i(x) + v_0(x)  e^{i\frac{\theta_i}{2}+i\frac{\pi}{4}} \psi^\dagger_i(x)\},
~
\eea
where $\theta_i$ is the Cooper pairing phase in the $i$-th chain.
We denote $\Delta$ the magnitude of the bulk gap,
$E_f$ and $k_f$ are the Fermi energy and Fermi momentum, respectively.
In the case of $\Delta \ll E_f$, the zero mode wavefunction is
approximated as
\bea
u_0(x)=v_0(x)\approx e^{-\frac{L-x}{\xi_x}}\sin k_f x,
\eea
where  $\xi_x=\frac{2}{k_f}\frac{E_f}{\Delta}$ is the coherence length
along the $x$-direction.
Now let us connect two right ends with a weak link
\bea
H_J= -t_\perp \int^{L}_{L-w} dx ~[\psi^\dagger_1(x) \psi_2(x)+h.c.],
\eea
where $w$ is the width of the link.
We assume $w \ll \xi_x$ such that the uniform distribution of the
pairing phase within each wire is not affected by the link.
By expressing $\psi_i(x)= u_0(x) e^{i\frac{\theta_i}{2}+i\frac{\pi}{4}}
\gamma_{i}+...$ where ``...'' represents for Bogoliubov eigenstates
outside the gap $\Delta$, we arrive at two different contributions for the
Josephson couplings: the usual one through second-order perturbation
process, and the fractionalized one through the zero Majorana mode as
\bea
E_J=-J_0 \cos \Delta \theta -  i J_1\gamma_{1}\gamma_{2}
\sin\frac{\Delta\theta}{2}
\label{eq:josephson}
\eea
where $\Delta\theta =\theta_1-\theta_2$,
$J\propto t_\perp^2/\Delta$, and $J_1=t_\perp \int^{L_x}_{L_x-\Delta L}
dx u_0(x) u_0(x+\Delta y)$.

\begin{figure}[tbp]
\centering\epsfig{file=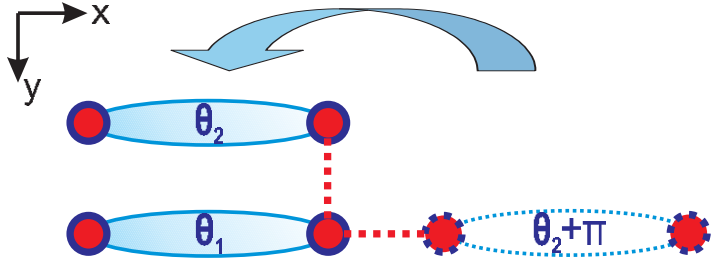,clip=2,width=0.6\linewidth, angle=0}
\caption{The fractionalized Josephson coupling between two parallel wires
through the Majorana zero modes depends on $\sin\frac{\Delta\theta}{2}$
instead of $\cos\frac{\Delta\theta}{2}$.
Please note the different orientations of the $x$ and $y$-directions
between Fig. \ref{fig:current} and this figure.
}
\label{fig:two-wires}
\end{figure}

The fractionalized Josephson coupling of the weak link between two
wires with the ``head-to-tail'' configuration \cite{kwon2004, xu2010} gives rise to
$i\gamma_1\gamma_2\cos\frac{\Delta \theta}{2}$,
where $\gamma_{1,2}$ are Majorana modes at two sides of the link.
In contrast, for two parallel wires, the corresponding
$J_1$ term in Eq. \ref{eq:josephson} is proportional to
$\sin\frac{\Delta\theta}{2}$.
This difference can be intuitively understood as
illustrated in Fig. \ref{fig:two-wires}.
We rotate the 2nd wire by $180^\circ$, which corresponds to a phase shift
of the 2nd wire by $\pi$ due to the $p_x$ symmetry, and
thus $\Delta\theta$ is shifted to $\Delta\theta +\pi$.
This changes the dependence of fractionalized Josephson coupling from
cosine to sine in Eq. \ref{eq:josephson}.
As a result, the Majorana fermion bonding strength vanishes at
$\Delta\theta=0$ which reflects the fact that these Majorana zero modes
are protected by the TR symmetry in the orbital channel.
The bonding strength varies linearly with the phase difference
as $\Delta\theta\rightarrow 0$, thus
Eq. \ref{eq:josephson} is minimized at nonzero values of $\Delta\theta_0$
satisfying $\sin\frac{\Delta\theta_0}{2}=\pm\min(\frac{J_1}{4J_0},1)$.
The energy is gained from the bonding of Majorana fermions.
For the general case that $\Delta\theta_0\neq \pm \pi$, 
the orbital channel time-reversal symmetry is spontaneously broken.
However, in the case of $\frac{J_1}{4J_0}\ge 1$, $\Delta\theta_0=\pm \pi$,
the Majorana fermion bonding strength is maximal.
Because of the $2\pi$ periodicity of phase angles, this case is also TR
invariant.

We can further extend Eq. \ref{eq:josephson} to a group of parallel wires
whose ends at $x=L$ are weakly connected, which
forms 1D or 2D lattices of Majorana modes coupled to the superfluid
phases.
The effective Hamiltonian is
\bea
H_{mj}&=&-J_0\sum_{\avg{ii^\prime}} \cos (\theta_i-\theta_{i^\prime})
-iJ_1 \sum_{\avg{ii^\prime}} \sin\frac{\theta_i-\theta_{i^\prime}}{2}
\gamma_i \gamma_{i^\prime},
\label{eq:maj_lattice}
\eea
where $\avg{ii^\prime}$ refers to the nearest neighbor bonds.
The $J_0$ term favors a global phase coherence with a constant phase
value in each wire.
However, this results in a flat band of zero energy Majorana fermions
which is highly unstable.

In bipartite lattices, a staggered phase configuration
$\theta_i=(-)^i \theta_0^\prime$ can minimize the ground state
energy by generating a uniform bonding strength, where
the value of $\theta_0^\prime$ can be obtained self-consistently.
For a 2D lattice of surface Majorana fermions, if the superfluid
phase distribution
of $\theta_i$ forms a vortex, the plaquette in which the vortex core
is located exhibit a $Z_2$ vortex for the Majorana fermion.
The motion of the superfluid vortex introduces the dynamic
$Z_2$ flux for the Majorana fermions.

\section{Spontaneous staggered orbital currents 
near the boundary}
\label{sect:boundary}
Having presented the spontaneous orbital channel time-reversal symmetry
breaking of Majorana fermions with weak links, now let us consider a 2D 
superconductor of a single component fermion with the $p_x$ pairing.
The bulk of the system with $p_x$-pairing maintains time-reversal symmetry 
in the orbital channel unlike the $p_x+ip_y$ one in which time-reversal
symmetry is broken.
We use the tight-binding model of the following mean-field Bogoliubov-de
Gennes (B-de G) Hamiltonian with the open boundary condition
\bea
H_{mf}&=&-\sum_{i} \Big\{ (t_x c^\dagger_{i} c_{i+\hat e_x}
+t_y c^\dagger_{i} c_{i+\hat e_y}+ h.c.)
-\mu c^\dagger_{i} c_{i} \Big\}\nn \\
&-&V\sum_{i} \Big\{ \Delta_{i,i+\hat e_x}^* c_{i+\hat e_x} c_i
+h.c.\Big\} 
+V\sum_{i}\Delta^*_{i,i+\hat e_x} \Delta_{i,i+\hat e_x},
\label{eq:bdeg}
\eea
where $\Delta_{i,i+\hat e_x} =\langle c_{i+\hat e_x}c_{i}\rangle$.
In conventional superconductors, the open boundary only affects the
magnitude of the pairing order parameter not on the phase distribution.
The $p_x$-type superconductors are weakly topological in 2D, and as is
well-known, if the phase distribution of $\Delta_{i,i+\hat e_x}$ is uniform such
that time-reversal symmetry is maintained in the oribtal channel, a 
flat band of Majorana surface modes appear on the boundary along 
the $y$-direction.

\begin{figure}[tbp]
\centering\epsfig{file=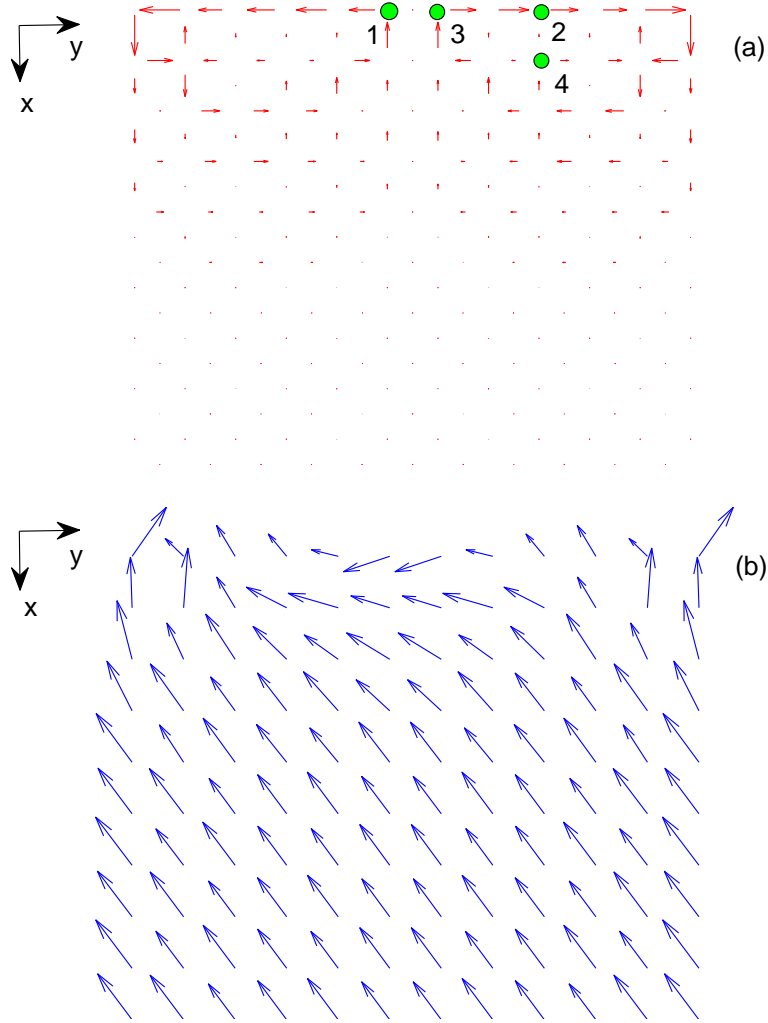,width=0.7\linewidth, angle=0}
\caption{The current and phase distributions from the self-consistent
solutions to Eq. \ref{eq:bdeg} with the open boundary condition.
The system size is $L_x\times L_y$ with $L_x=20$ and $L_y=12$, and
parameter values $t_x=1$, $t_y=1.2$, $V=3$, and $\mu=0$.
Only half of the system along the $x$-direction is depicted, and the
distributions in the rest half can be obtained by performing the
reflection operation.
a) The spontaneously generated edge current distributions;
b) the lengths and directions of arrows represent  magnitudes and
phase angles $\theta_{i,i+\hat e_x}$ for the pairing operators
$\Delta_{i,i+\hat e_x}$, respectively.
}
\label{fig:current}
\end{figure}

However, this Majorana band flatness is unstable when the
spatial variations of pairing phases are considered.
It spontaneously breaks time-reversal symmetry and generates finite bonding
strengths among the Majorana modes.
Consequently, the macroscopic degeneracy of the surface zero modes is
removed and the total energy is lowered.
To confirm this intuitive picture, we perform the self-consistent solutions
to the B-de G equation Eq. \ref{eq:bdeg} at zero temperature.
The mean-field Hamiltonian is diagonalized to obtain the Bogoliubov
eigenstates whose eigen-operators are defined through
$c_{i} =\sum_n u_{n}(i)\gamma _{n}+v_{n}^{\ast }(i)\gamma
_{n}^{\dagger }$,
where $(u_n, v_n)^T$ are the eigenvectors with positive energy eigenvalues
$E_n>0$.
The gap equations for $\Delta_{i,i+\hat e_x}$ read
\bea
\Delta_{i,i+\hat e_x}&=& \frac{1}{2}\sum_n
\tanh \frac{\beta E_n}{2} \left[u_{n}(i+\hat{x})v_{n}^{\ast }(i)
-v_{n}^{\ast }(i+\hat{x})u_{n}(i) \right].
\eea
The current along the bond between $i$ and
$i+\hat e_a$ $(a=x,y)$ is
\bea
j_{i,\hat e_a} =-\frac{t_a}{\hbar }\sum_n
\mbox{Im}\Big\{ v_{n}(i+\hat e_a)v_{n}^{\ast}(i) f(E_n) 
+u_{n}(i+\hat e_a)u_{n}^{\ast }(i)[1-f(E_n)] \Big\}, \nn \\
\label{eq:current}
\eea
where $f(E)$ is the Fermi distribution function.

\begin{figure}[tbp]
\centering\epsfig{file=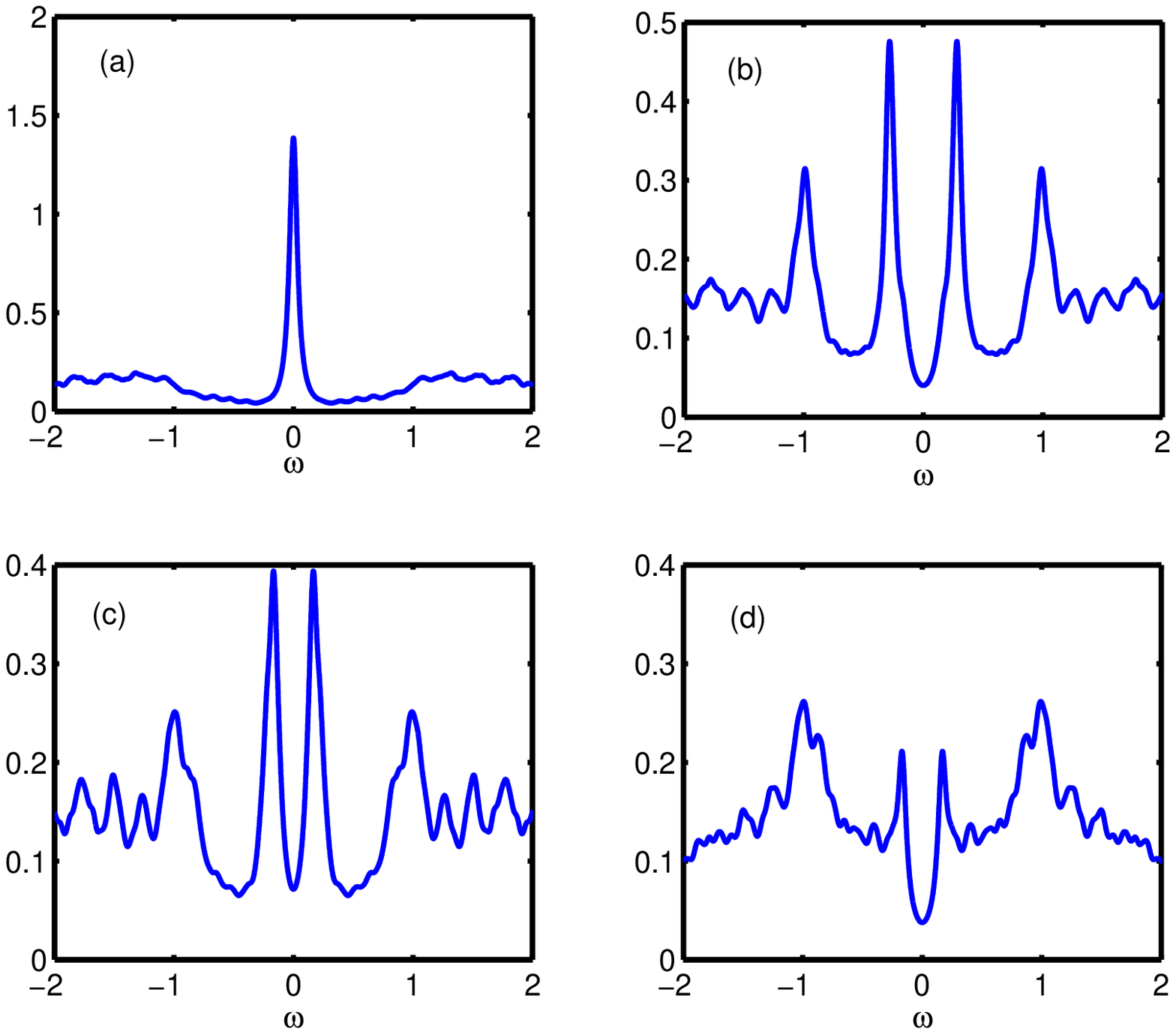,clip=2,width=0.8\linewidth, angle=0}
\caption{The edge LDOS spectra for the system in Fig. \ref{fig:current}
$\delta(\omega\pm E_n)$ in Eq. \ref{eq:LDOS}
are approximated by $\frac{1}{\pi}\eta/[(\omega\pm E_n)^2+\eta^2]$ with the
broadening parameter $\eta=0.04$.
a), b), c) and d) are the LDOS spectra at points 1, 2, 3, and 4 marked
in Fig. \ref{fig:current} a), respectively.
a) LDOS without self-consistency by fixing $\Delta_{i,i+\hat e_x}$ at
a uniform bulk value.
b), c) and d) are self-consistent results.
The corresponding parameter values are presented in the
caption of Fig. \ref{fig:current}.
}
\label{fig:LDOS}
\end{figure}

The spontaneous time-reversal symmetry breaking effect in the orbital
channel appears near the boundary
normal to the $x$-direction: circulation current loops as depicted in
Fig. \ref{fig:current} a).
As presented in Fig. \ref{fig:current} b), this edge induces non-uniform
phase distribution of the pairing order parameter $\Delta_{i,i+\hat e_x}$.
However, it is non-singular which does not exhibit the
vortex configuration and thus cannot give rise to the current loops
by itself.
Loosely speaking there are two different contributions
to the current expression of Eq. \ref{eq:current}: the Bogoliubov
modes whose energies outside the bulk gap,
and the in-gap states of Majorana modes.
The former contribution can be captured by the Ginzburg-Landau
formalism as spatial phase variations of $\Delta_{i,i+\hat e_x}$,
while the latter arises from  Majorana modes cannot.
The circulation pattern of these induced edge currents are staggered,
which is natural, since in the background of superfluidity the overall
vorticity should be neutral in the absence of external magnetic fields.
The size of current loops should be at the order
of coherence lengths which can be estimated as $\xi_a/a_0\approx 2t_a/\Delta$
with $a=x,y$ where $a_0$ is the lattice constant. 
Currents decay exponentially along the $x$-direction at the scale
of $\xi_x$, and change directions along the $y$-direction at the
scale of $2\xi_y$.

Next we calculate the local density of states (LDOS) near the boundary
as depicted in Fig. \ref{fig:LDOS}.
The expression for LDOS reads
\bea
L(i,\omega)=\sum_n |u_n(i)|^2 \delta(\omega-E_n)
+|v_n(i)|^2 \delta(\omega+E_n).
\label{eq:LDOS}
\eea
For comparison, we first present the result without self-consistency in
Fig. \ref{fig:LDOS} a) by fixing order parameters $\Delta_{i,i+\hat e_x}$
uniform with the bulk value.
The central peak at zero energy represents the edge Majorana states.
However, with self-consistency, spatial variations of $\Delta_{i,i+\hat e_x}$
couple to the edge Majorana modes.
The zero energy peaks are split as depicted in Fig. \ref{fig:LDOS} b), c)
and d).
Sites 2, 3 (Fig. \ref{fig:LDOS} b) and c), respectively) are right on the
boundary normal to $x$-direction, and thus their coherence peaks are
strongly suppressed.
Site 4 (Fig. \ref{fig:LDOS} d) is relatively inside, and thus the midgap
peaks are suppressed.
Site 2 locates in the middle of current loops, and thus the splitting
of the zero bias peak is the largest.

Due to the divergence of density of states at zero energy, the above 
time-reversal symmetry breaking mechanism in the orbital channal
is general.
It also applies to the 3D case with the $p_z$-pairing.
Then the $xy$-surface is non-trivial along which Majorana
zero energy flat bands appear with the assumption of time-reversal symmetry.
Again due to the same reasoning, this degeneracy will be lifted by
spontaneous TR symmetry breaking.
In this case, we expect that the 2D vortex-anti-vortex pattern in Fig.
\ref{fig:current} will change to that of 3D closed vortex rings,
which may further form a lattice structure parallel to the $xy$-surface.
Further study of this lattice structure will be deferred to a later
publication.
Moreover, the above physics also applies in the continuum not just
in the tight-binding model.
Staggered vortices or closed vortex rings will appear near the edge
or surfaces within the size of coherence lengths, which will also
split the zero energy peaks of Majorana fermions.

Next we briefly discuss the boundary Majorana flat bands for the TR
invariant spinful $p_x$-wave Cooper pairing.
For example, let us consider the pairing order parameter in the bulk
as $\Delta_{i,i+e_x}=i |\Delta|\sigma_2 \vec \sigma \cdot \vec d $.
Without loss of generality, we can assume that the $d$-vector
lies in the $xy$-plane with the azimuthal
angle $\phi$, i.e., $d_x+id_y=|d|e^{i\phi}$.
This corresponds to that spin-up fermions are paired with spin-up,
and spin-down ones are paired with spin-down, with the
relation $\Delta_{\uparrow\uparrow}=|\Delta|e^{i\phi}$
and $\Delta_{\downarrow\downarrow}=|\Delta|e^{-i\phi}$.
Then the boundary Majorana modes consist two branches
$\gamma_\uparrow$ and $\gamma_\downarrow$ forming Kramer pairs.
The bonding among Majorana modes can occur within each branch of Kramer
pair by developing currents as explained above.
If a residue magnetic interaction exists, another way to remove
the degeneracy is to develop
magnetism with the order parameter $i\gamma_\uparrow \gamma_\downarrow$.
The spin polarization for such an order is in the $xy$-plane.
In general, these two different bonding mechanisms can coexist.

\section{Discussions}
\label{sect:discussions}

The above real $p$-wave Cooper pairing of single component fermion
can be realized in ultra-cold dipolar fermionic molecular systems.
For example,  the $^{40}$K-$^{87}$Rb systems have been cooled down
below quantum degeneracy \cite{ni2010,ospelkaus2010}.
Moreover, the chemically stable dipolar fermion molecules of
$^{23}$Na-$^{40}$K have also been laser-cooled \cite{wuch2012}.
If the dipole moments are aligned along the $z$-axis by external
electric fields, the interaction between two dipole moments exhibits
the $d_{r^2-3z^2}$-type anisotropy.
It means that the interaction is attractive if the displacement vector
between two dipoles is along the $z$-axis, and repulsive if it lies
in the $xy$-plane.
Naturally, this leads to the $p_z$-type Cooper pairing in the
strong coupling limit.
Even in the weak coupling limit, partial wave analysis also 
shows that the dominant pairing symmetry is the $p_z$-type.
Due to the anisotropy of the interaction, it slightly
hybridizes with other odd partial wave channels as predicted in
previous works \cite{baranov2002,baranov2004,you1999}.
For the two-component dipolar fermions, the triplet $p_z$ Cooper
pairing has also been theoretically predicted \cite{wu2010,shi2010}.
The above predicted spontaneous time-reversal symmetry breaking effects
in the orbital channel will appear on the surfaces perpendicular to the $z$-axis.

In the condensed matter systems, the boundary Majorana fermion has 
been detected in the 1D quantum wire with SO coupling
\cite{mourik2012,churchill2013,das2012,finck2013,rokhinson2012}.
A magnetic field is also applied in parallel to the wire to lift the
degeneracy at zero momentum such that the system is effectively
one component if the Fermi energy lies inside the gap.
In this case, the major effect of the magnetic field is the
Zeeman effect, and the time-reversal symmetry is not explicitly 
broken in the orbital channel. 
The proximity effect from the superconducting electrodes induces
superconductivity in the 1D wire, which leads to the zero energy
Majorana ending states.
If we put an array of these 1D wires together and their ends are
weakly connected, these Majorana modes will also spontaneously
build up bonding strength by developing staggered orbital currents
patterns, the consequential splitting of the zero bias peaks will be
exhibited in the tunneling spectroscopy.

\section{Conclusion}
\label{sect:conclusion}
In summary, we have found that the zero energy boundary Majorana
flat bands of the $p$-wave superconductors
are unstable towards spontaneous time-reversal symmetry breaking
in the orbital channel,
which results from the coupling between Majorana modes and the
Cooper pairing phases.
The divergence of density of states at zero energy leads to
spontaneous formation of staggered current loops, which generates
bondings among the Majorana modes and lifts the degeneracy.
This is a robust mechanism which appears even without introducing
new instabilities.

\ack{We thank Z. Cai for early collaboration on this work, and
L. Fu, B. Halperin, and P. A. Lee for helpful discussions.
Y. L., D. W., and C. W. are supported by the NSF DMR-1105945 and
AFOSR FA9550-11-1- 0067(YIP);
Y. L. is also supported by the Inamori Fellowship.}

{\it Note added}
Part of this work was presented in the 2011 APS march meeting \cite{li2011}.
Near the completion of this manuscript, we learned the nice work
\cite{potter2013}, which studies the magnetic instabilities 
in Majorana flat bands.

\section*{References}


\end{document}